\documentclass[journal]{IEEEtran}
\usepackage{amsmath,amssymb,amsmath,amsfonts}
\usepackage{bm}
\usepackage{bbm}
\usepackage{graphicx}
\usepackage{cite}
\usepackage{epstopdf}
\usepackage{balance}
\usepackage{gensymb}
\usepackage{color}
\usepackage{lipsum}
\usepackage{float}
\usepackage{epstopdf}
\usepackage[mathcal]{eucal}
\usepackage{subfiles}
\usepackage[T1]{fontenc}
\usepackage{siunitx}
\usepackage{tabulary}

\usepackage{epsfig,graphics,graphicx,amssymb,amstext,amsmath,algorithm,algorithmic,wrapfig,multirow}
\usepackage{array}
\usepackage{adjustbox}
\usepackage[dvipsnames]{xcolor}
\usepackage{cite}
\usepackage{times}
\usepackage{placeins}
\usepackage{textcomp}
\usepackage[font=small]{caption}
\usepackage[breaklinks=true]{hyperref}
\usepackage{xcolor}
\usepackage{tabularx}
\usepackage{bm}
\usepackage{enumerate}
\usepackage{tikz}
\usepackage{pgfplots}
\usepackage{epstopdf}
\usetikzlibrary{shapes,arrows}
\usepackage[utf8]{inputenc}
\usepackage{mathtools}
\usepackage[utf8]{inputenc}
\usepackage{xcolor}
\usepackage{tikz}
\usepackage{multirow}
\usetikzlibrary{chains,arrows,calc,positioning}
\usepackage{amssymb}% http://ctan.org/pkg/amssymb
\usepackage{pifont}% http://ctan.org/pkg/pifont
\usepackage{caption}
\usepackage{subcaption}

\pgfplotsset{compat=1.16}
\definecolor{bittersweet}{rgb}{1.0, 0.44, 0.37}
\definecolor{glaucous}{rgb}{0.38, 0.51, 0.71}
\definecolor{gainsboro}{rgb}{0.86, 0.86, 0.86}
\definecolor{babyblueeyes}{rgb}{0.63, 0.79, 0.95}
\definecolor{silver}{rgb}{0.75, 0.75, 0.75}
\definecolor{neoncarrot}{rgb}{1.0, 0.64, 0.26}

\usepackage{soul}

\usepackage{breqn}
\usepackage{booktabs}
\usepackage{multirow}

\usepackage{enumitem}
\usepackage{tabulary}

\usepackage[normalem]{ulem}
\makeatletter

 \let\oldforeign@language\foreign@language
 \DeclareRobustCommand{\foreign@language}[1]{%
   \lowercase{\oldforeign@language{#1}}}

\usepackage{dblfloatfix}
\usepackage{makecell}

\makeatother

\IEEEoverridecommandlockouts

\begin{document}
\title{THz-Empowered UAVs in 6G: \\
Opportunities, Challenges, and Trade-Offs}
% A Tutorial on 
% What's [...] ?

\author{M. Mahdi Azari,~\IEEEmembership{Member,~IEEE}, Sourabh Solanki,~\IEEEmembership{Member,~IEEE},\\ Symeon Chatzinotas,~\IEEEmembership{Senior Member,~IEEE}, Mehdi Bennis,~\IEEEmembership{Fellow,~IEEE}
\thanks{M.~M.~Azari, S.~Solanki, S.~Chatzinotas are with the SnT, University of Luxembourg. M. Bennis is with the  University of Oulu.}
%\thanks{M. Bennis is with the  University of Oulu.}
}

\maketitle

\begin{abstract} 
 Envisioned use cases of unmanned aerial vehicles (UAVs) impose new service requirements in terms of data rate, latency, and sensing accuracy, to name a few. %Increasing demands for the deployment of unmanned aerial vehicles (UAVs) in new applications, particularly envisioned in 6G, impose new service requirements in terms of, data rate, latency, and sensing accuracy, to name a few. 
 If such requirements are satisfactorily met, it can create novel applications and enable highly reliable and harmonized integration of UAVs in  the 6G network ecosystem.
%Such requirements, if satisfied properly, even create new ambitious applications and enable highly reliable and harmonized integration of UAVs into 6G. 
 Towards this, terahertz (THz) bands are perceived as a prospective technological enabler for various improved functionalities such as ultra-high throughput %due to abundant bandwidth 
 and enhanced sensing capabilities. %due to shorter wavelengths.
%Relevant to 6G, terahertz (THz) bands are prospective technological enabler for various improved functionalities such as higher throughput due to more available bandwidth and sensing capabilities due to shorter wavelengths. 
This paper focuses on THz-empowered UAVs with the following capabilities: communication, sensing, localization, imaging, and control. We review the potential opportunities and use cases of THz-empowered UAVs, corresponding novel design challenges, and resulting trade-offs. Furthermore, we overview recent advances in UAV deployments regulations, THz standardization, and health aspects related to THz bands. Finally, we take UAV to UAV (U2U) communication as a case-study to provide numerical insights into the impact of various system design parameters and environment factors. %A comparative study shows that although the impact of path loss and molecular noise is significantly higher in THz bands, THz-empowered U2U communication offers higher achievable throughput compared to mmWave and sub-6 GHz up to several hundred meters.
\end{abstract}
%  \begin{IEEEkeywords}
% Unmanned aerial vehicle (UAV), terahertz (THz), 6G, mmWave, sub-6 GHz.
% %\vspace{-0.3 cm}
%  \end{IEEEkeywords}
%\vspace{-0.1in}
\section{Introduction}

%Inherent attributes of unmanned aerial vehicles (UAVs) such as their mobility and flexibility in 3D positioning combined with the technological advancement in cost-effective manufacturing of UAVs, motivate the delegation of more tasks to UAVs in upcoming wireless networks. 
In the envisioned 6G networks, UAVs are going to play a prominent role in several aspects such as network intelligence and autonomy \cite{azari2021evolution} thanks to their flexibility in 3D positioning and their technological advancement in cost-effective manufacturing. On the other hand, to satisfy the proliferating demands in 6G, terahertz (THz) band stands among the candidate enablers \cite{azari2021evolution, giordani2020toward}. This is owing to the underlying potential of THz band in characterizing higher throughput, lower latency, accurate localization, and precise sensing and imaging capabilities. Recent advancements in the field of semiconductor industries enable design of more compact THz devices %for signal generation and radiation, 
which was traditionally one of the major barriers in THz deployment \cite{cYiJSAC2021}. Therefore, towards the progression of 6G networks, THz-empowered UAVs may offer several opportunities, opening the door for new services and novel applications \cite{azari2021evolution}. Regarding this, Section \ref{sec:opportunities} presents the prospective use cases and opportunities enabled by THz-empowered UAVs. 

Despite the huge potential of THz band, there is still a long way to go before we see its practical applications in wireless communications. However, the persistent efforts across the research community over the past few years helped this speculative technology become increasingly attainable. The roadmap towards advancing the technology further has also been actively discussed in various scenarios \cite{akyildiz2014terahertz, rappaport2019wireless}. %The use of THz frequencies, however, is not well investigated yet for the ground and near ground deployments. Accordingly, there are unknown features to be discovered or verified. 
THz-empowered UAVs not only face the challenges coming from the nature of such frequencies, they are also affected by the intricacies of efficient UAVs deployment. For example, molecular noise, absorption loss, and uncertain medium are major issues with the THz transmissions which become more aggravating in outdoor scenarios with UAVs. On the other hand, UAVs flexibility of positioning them in line-of-sight (LoS) and adjusting their distances have great potential in using THz bands effectively. This paper, in Section \ref{sec:challenges}, highlights important design challenges of THz-empowered UAVs along with induced trade-offs to provide relevant insights into the new system design.

Certain barriers in the deployment of THz-empowered UAVs include regulatory and health aspects pertaining to the use of THz frequencies, which are overlooked in most academic reports. Such challenges are elaborated in Section \ref{sec:regulations}, where the advancement of regulations, standardization, and safety aspects are reviewed. This section is followed by Section \ref{sec:case_study} where several system design parameters and environment factors for quality communication between UAVs are recognized and their impacts are investigated.A few works have analyzed the performance of THz communications for UAV-ground links \cite{wangTVT,rajaInfocom}, but the underlying intricacies of combining the two technologies were not discussed. To our best knowledge, this work is the first one that provides a comprehensive study on UAVs empowered by THz functionalities where potential use cases, new challenges, design trade-offs, regulatory and health aspects, as well as numerical insights are discussed. 
%\vspace{-0.1in}

% \begin{itemize}
%     \item motivation
%     \item related works
%     \cite{azari2021evolution}
%     \item contribution
%     \item paper structure
% \end{itemize}

\section{Fundamental Features of Terahertz Communications} \label{sec:fundamentals}
THz propagation exhibits certain distinct attributes reviewed in the sequel.%in contrast to the microwave/mmWave band. Though some characteristics can be similar as that of mmWave, they are more conspicuous and prominent at THz band.

\subsubsection{Channel and Propagation Characteristics}
Unlike lower frequencies, wavelengths of THz are comparable to the size of rain, dust, or snow, which makes attenuation due to molecular absorption more severe. Specifically, the atmospheric absorption due to oxygen ($\textrm{O}_2$) and water ($\textrm{H}_2$O) molecules is significant in the THz channel. Since this absorption is dependent on the concentration and composition of the molecules along the path of the waves, the losses can vary based on the meteorological conditions thereby making THz band highly frequency-selective. THz waves also encounter spreading loss due to expansion of EM waves along the medium. The path loss due to absorption and spreading account for the LoS propagation of the THz waves. The path loss becomes more significant with increase in the transmission distance, however, the molecular absorption loss can be minimized by selecting a suitable bandwidth (0.38-0.44 THz, for example) where the losses are well below 10 dB per km \cite{akyildiz2014terahertz}. 
In addition to aggravating the attenuation, molecular absorption also introduces a non-white noise depending on the mixture of molecules.

On the other hand, when LoS propagation is not feasible due to obstacles, THz waves undergo diffuse scattering, specular reflections, diffraction etc., which account for the non-LoS (NLoS) propagation. These characteristics are dependent on the geometry and material of the impinging surface as well as the incident angle of EM wave. 
%However, various works often ignore the NLoS propagation due to dominant LoS path. %\footnote{\mahdi{double check the message you d like to convey!\sourabh{checked}}} 
%The wide-band bandwidth at THz leads to varying attenuation and delay with reference to each frequency components which eventually requires new multipath channel models.
Essentially, the THz channel models should encapsulate the free space and absorption losses which are dependent on frequency, distance, altitudes, and relative air composition. Also, the path loss is a function of 3D LoS/NLoS probability. %Additionally, multipath fading is an important aspect too for the channel modelling.
Nonetheless, channel modelling for THz propagation is yet to be fully understood. Several measurement campaigns and  research efforts are ongoing for its characterization \cite{rappaport2019wireless}.%. A more detailed picture of the propagation modelling for the THz band can be found in \cite{HanMag18}.}
%In particular, the THz waves are susceptible to various losses such as atmospheric loss, diffuse scattering, specular reflections, diffraction, shadowing, scintillation effects, etc. 
%Therefore, traditionally existing channel models available for the traditional RF transmission may not be directly applicable for THz waves and new accurate channel models need to be developed which can encompass all the propagation characteristics. 
%\subsubsection{Molecular and Atmospheric Absorption}

\subsubsection{Large Antenna Arrays for Pencil Beamforming} 
To mitigate the high transmission loss at THz, the adoption of large antenna arrays for directional beamforming is inevitable. In fact, extremely small wavelengths can facilitate the small-sized antenna design that can be deployed on a large scale to form huge antenna arrays for sharped pencil beams. Such directional transmission can compensate for the attenuation up to a certain extent. However, it brings about additional challenges such as mobility and handover management and efficient beam tracking and alignment. Further, wideband beamforming is also a critical problem primarily caused by spatial-wideband effect due to the use of large antenna arrays and frequency-selectivity due to multipath delay \cite{gaojsac}. Wideband effects are also prevalent at mmWave band but they are more prominent at THz and require extensive research efforts to explore the appropriate solutions. 

\begin{figure}
\centering 
\includegraphics[width=\columnwidth]{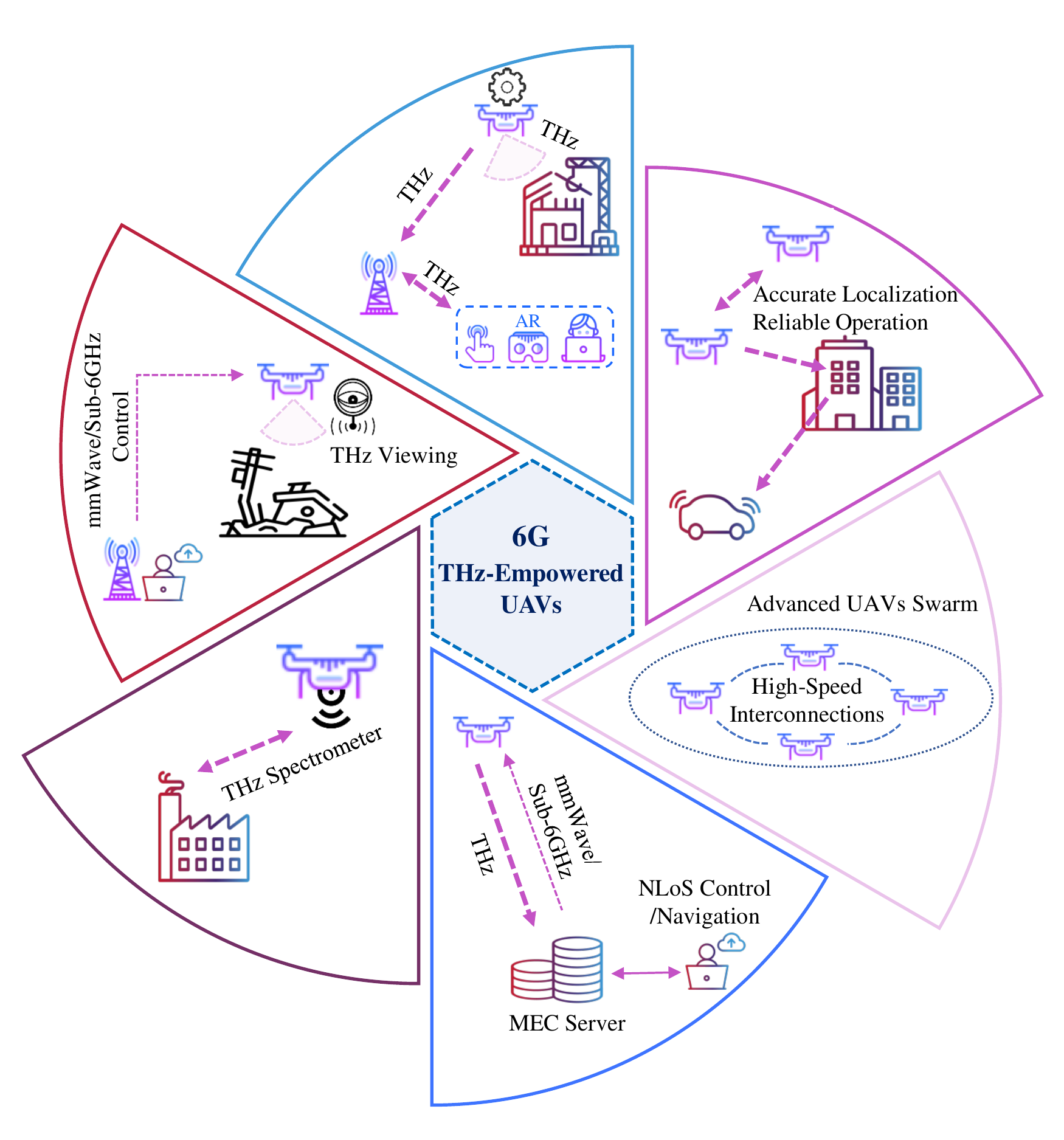}
\caption{Representative use cases of THz-empowered UAVs.}
\label{fig:use_cases}
\end{figure}

%\vspace{-0.1in}
\section{Selected Use Cases and Opportunities} \label{sec:opportunities}
In the following, we discuss several important applications of THz-empowered UAVs as illustrated in Figure \ref{fig:use_cases}.
\subsubsection{Safe and Reliable Operation of UAVs} % \red{[Sourabh]}
Large antenna arrays in small sizes at THz can enable cm-level localization accuracy and sensing solutions with finer range Doppler and angular resolutions \cite{bourdoux20206g}. Thus, THz-empowered UAVs possess the capability of accurate localization and sensing which can facilitate the instantaneous perception of the environment. Such functionalities lead to a safe and reliable navigation of THz-empowered UAVs. Note that the safe operation of UAVs is crucial for complying with the regulations in air-space and also for their practical application. Empowering UAVs with THz can help in achieving this objective. For instance, leveraging accurate sensing and localization, THz-empowered UAVs can flexibly position them with the intended transmitters/receivers to improve the reliability and safety. Moreover, by detecting the location and presence of any obstacles or other aerial nodes, they can be safely navigated while seamlessly avoiding any conflict or accident.
%Additionally, THz-UAVs can also assist to \textit{extract detailed 3D map} of an environment through beam scanning.
% \mahdi{elaborate a bit more this item. One reviewer asked to clarify the use cases better. remove the last sentence bc of space and consistency but explain it implicitly within this application.}
%\footnote{\mahdi{Can be replaced with Autonomous flight}}
    
%     In general, intelligent context-aware 6G use cases can exploit accurate localization and high resolution sensing information to optimize the deployments and energy usages. "Exploiting bandwidth and large antenna
% arrays for sensing: • Sensing the location and
% orientation of the user
% instantaneously (in the
% centimeter range).
% • High-resolution
% 3D mapping of the
% environment.
% • THz radar and
% object detection." \cite{chaccour2021seven}

\subsubsection{Interactive Aerial Telepresence} 
%To do unsafe, costly, or time-critical tasks for humans, UAVs can be deployed and controlled remotely while having humans in the loop to establish the so-called \textit{aerial telepresence}. The concept can be further enriched when combined with augmented reality which provides 3D visual feedback and real-time teleinteraction with the target environment. Such haptic guidance, indeed, enhances the UAV capability, creates new applications, and more importantly brings experts to the scene from anywhere at anytime. For this application scenario, particularly for the real-time interactions where the AR can not be compressed \cite{giordani2020toward}, the data rate may reach the Gbps and hence THz deployment plays an important role.
It is possible to deploy UAVs and control them remotely to perform unsafe, costly, or time-sensitive tasks for humans. This is referred to as \textit{aerial telepresence} and it becomes even more powerful when augmented reality (AR) is used which can offer 3D visuals and real-time tele-interaction with the environment. In fact, such haptic guidance enhances UAV capability, opens up new applications and, perhaps most importantly, provides access to experts from anywhere at any time. For such applications, THz deployment is essential particularly for real-time interactions where AR data cannot be compressed \cite{giordani2020toward} and high data rates are desired.
 For highly precise \textit{interactive tasks and immersive experience}, UAVs with THz sensors can provide accurate instantaneous perception of the environment thereby offering remote cognitive capabilities.
%Further to the communication rate and latency requirement, UAVs THz sensors can offer highly accurate environmental cognition and instantaneous perception of the environment being crucial for precision \textit{interactive tasks and immersive experience}.\footnote{\mahdi{this paragraph should be paraphrased.}}

\subsubsection{Advanced Swarm of UAVs} % \red{[Mahdi]}
To enhance the performance of UAVs networks and tackle the limits of an individual UAV, UAVs may perform in a swarm \cite{ngaojsac}. In these scenarios, 
%To enhance the performance of UAVs and tackle the limits of UAV, they may perform in \blue{ swarm \cite{ngaojsac} }by engaging in the same or different synergic activities. UAV swarm can be exploited for internet connectivity or they may perform as distributed aerial sensors to support specific tasks or enable full autonomy. UAVs in swarm may act as distributed computing or learning devices to overcome the limited on-board computing and energy capacity and also benefit from holistic data integration. In these applications, 
UAVs may need to exchange large amount of data, particularly given the significant growth of 6G data demand and increasing number of sensors per UAV for integrated services in 6G e.g., sensing, localization, mapping, etc. Accordingly, UAVs need to boost their interconnection rate capacity beyond the current technology, compatible with the upcoming generations and services. Ultra-high-throughput U2U communication link can be provided in THz given the favorable LoS dominant propagation and the fact that UAVs can establish sufficiently close communication range to mitigate high path loss and molecular absorption. %\mahdi{shorten this item.} %To maintain a   \mahdi{add distributed sensing, computing, learning using swarms. Distributed learning => to share the local models or data for a generalization of the algorithm/parameters. holistic data integration available in distributed nodes. Freshness of data can be enhanced using a high rate communication capacity. "To be driven by full autonomy, such systems need to exchange large amounts of data such as high-resolution real-time maps, with their environment, e.g., other vehicles, or BSs."} look at \cite[Sec. X]{chaccour2021seven} 
    
\subsubsection{Time-Critical MEC-Empowered Operation}  %\red{[Mahdi]} %\footnote{\sourabh{Most of the para talks about UAV and MEC. I suggest to put more emphasis on THz-UAV.}} \mahdi{shorten this item.}
Due to limited on-board computation capacity, UAVs may rely on mobile edge computing (MEC) servers which impose significant delay in obtaining required information. Providing sufficiently high-speed data link improves the  time-critical remote computations capability, enhances the freshness of data, and eventually enables instantaneous environmental cognition. MEC-empowered operation through THz links allows to reduce the UAVs payload resulting in lower propulsion energy consumption and longer operational lifetime. A real-time remote processing in turn creates new means of \textit{online NLoS UAVs control and navigation} through instantaneous perception of the environment. In this application, when the environment is unknown, three steps of sensing, processing, and navigating can be performed in real-time thanks to ultra-high-throughput THz links between UAVs and the computing-and-controlling servers. %Other examples include time-critical traffic control, surveillance, and autonomous UAVs operations.

\subsubsection{Pollution Monitoring and Tracking}  %\red{[Sourabh]}
 Frequency scanning spectroscopy at THz band allows air quality monitoring and detection of some chemicals since various materials at certain frequencies have vibrational absorption. The specific absorption characteristics of different gaseous medium significantly boost the THz's sensing capability. THz-spectroscopy coupled with UAV's flexibility can offer effective solutions for pollution monitoring and tracking. For instance, UAVs equipped with THz spectrometer can fly near the source of pollution (e.g., industrial chimneys) to detect the concentration of various pollutants. Also, THz-empowered UAVs can fly up to the mountain peaks to detect the water vapour concentrations. Such utilities of THz-empowered UAVs can help in combating the air pollution and assist to keep track of climate change.

\subsubsection{Rescue and Surveillance} %\red{[Mahdi]}
THz characteristics combine the advantages of microwaves and visible light for a practical implementation of NLoS imaging system. They involve large bandwidths and small wavelengths for high spatial resolutions, and also moderate scattering features for image reconstruction algorithms that are computationally less expensive than optical systems \cite{rappaport2019wireless}. Although the range of THz NLoS imaging might be yet short, UAVs flexibility in 3D portioning can address such limit by hovering in the proximity of a target area. UAVs equipped with THz waves can view NLoS objects during \textit{rescue and surveillance}, expanding the benefits of THz imaging systems. %\footnote{\mahdi{update this item based on 2 characteristics: 1- THz sensors, 2- mobile THz sensors acting in BVLoS and NLoS. and maybe add it to the figure}. \sourabh{Can this be the part of Aerial Telepresence item?} } %\footnote{\mahdi{what is  the size of imaging system? can it be mounted on UAV, at least part of it that is needed.}}
\section{Design Challenges and Trade-Offs} \label{sec:challenges}

% THz Communication encounters several challenges related to the technology itself. For example, the noise power not only increases with the larger bandwidth but also accompanied by molecular absorption noise. Indeed, the molecular absorption noise intensify the overall noise in addition to its detrimental effect on the received signal power. 
%The absorption coefficient depends highly on the frequency such that there are peaks of significantly high atmospheric absorption. Such peaks should be avoided when planning the use of THz frequencies. Fortunately, as mentioned in Section \ref{sec:fundamentals}, there are abundant GHz windows where the molecular absorption is reasonable and relatively low. %, i.e. below 10 dB/Km \cite{akyildiz2014terahertz}. 
% Next, we list specific challenges in the deployment of THz-empowered UAVs for different services. 

In the following, we list specific challenges and potential trade-offs in the deployment of THz-empowered UAVs for different services. 
\subsection{Challenges}
\subsubsection{UAVs Wobbling and Motion}
Effective use of THz frequencies requires highly directional antennas at both transmitter and receiver sides to compensate the large propagation loss. However, UAVs wobbling/fluctuation due to wind can result in beam misalignment which deteriorates the communication quality. Also, due to uncontrolled tilts and rotations from UAVs motion, maintaining a perfect beam alignment would be intricate thus, resulting into several beam hopping and handover issues. Such impact gets worse as antenna footprint shrinks with larger antenna gain. %indicating a natural trade-off between higher antenna gain and frequent beam misalignment.
%Moreover, UAV's mobility causes several beam hopping and handover issues due to pencil beams. 
Even if such phenomena does not heavily reduce the overall rate, more frequent handovers results in significant delay and adversely impact the link reliability. 
To guarantee a certain level of reliability under delay constraint, some handovers may be skipped at the expense of rate reduction % which highlights an important underlying trade-off. 
Further, we recall that sensing and localization capability in THz can assist UAVs to operate reliably by precisely positioning the intended transmitters/receivers.
In addition, intelligent reflective surface (IRS) enabled THz architecture can ease effective beam alignment by offering enhanced sensing and localization \cite{chaccour2021seven}.
%\sourabh{Please check if some important information is missed.}
%Further, we recall that accurate 3D sensing and localization capability in THz can assist UAVs to operate safely and reliably by precisely positioning the intended transmitters/receivers.
\subsubsection{Power-Hungry Payload} % \red{[Mahdi]}
High hardware power consumption in THz, essentially corresponding to the analog-to-digital converters (ADCs), is one of the main challenges for UAVs with limited power budget. Although ultra-wide bandwidths in THz band appears to be the major motivation to move towards higher frequencies, the power consumption of ADCs remarkably increases when sampling rate surpasses 100 MHz which grows exponentially with resolution  \cite{Murmann}. %Therefore, a new visit to the design of energy-efficient THz-empowered UAVs is required. 
At lower frequencies, it is widely accepted that the propulsion power consumption is dominant over the electrical part which facilitates several energy-efficient designs. In THz, given the high number of antennas with the corresponding RF chains and ADCs the total payload power consumption may not be negligible as compared to that of propulsion. Therefore, it is important to update the design guidelines for THz-empowered UAVs. To plan the THz-empowered UAV trajectory, longer traveling is needed for a meaningful data transfer in contrast to low frequencies.  %it may be more energy efficient if the flight is longer in order to establish a more favorable LoS link with a destination instead of making traveling lines shorter for lower propulsion energy consumption. %Moreover, the capacity of communication link in THz remarkably changes with distance and hence in general longer traveling might be needed for a meaningful data transfer in contrast to low frequencies. This might be more energy consuming, however, the amount of transferred data could also be significantly higher. 
% \blue{Accordingly, there are more complex trade-offs between increased energy consumption due to additional payload-related power consumption and the longer flying time, and the higher possible data rate in THz.} %In Section \ref{sec:case_study}, we compare distance-dependent link throughput at different frequency bands. 

%\mahdi{mention the payload weight challenges here.}%\footnote{\mahdi{the high power consumption needed for sensing and NLoS radars.} \mahdi{ figure/table to compare the power consumption of different hardware units (maybe in both Sub-6GHz, mmWave and THz) and the propulsion energy consumption => look at the refs of the cited papers.} \mahdi{ "Toward Millimeter Wave Joint Radar-Communications: A Signal Processing Perspective"}}

\subsubsection{Limited Design Space} 
%The importance of optimal 3D placement of UAVs in sub-6 GHz have been recognized in several studies. 
%It has been shown that, in several application scenarios such as UAV acting as aerial BS there is an optimal UAV altitude for maximal coverage, rate, or reliability where transition from NLoS to LoS and hence the more favorable propagation condition compensates the destructive impact of longer link length at altitudes. 
%For example, in applications where UAV acts as an aerial BS, an optimal UAV altitude is desirable for maximum coverage, rate, and reliability. 
Optimal height of UAVs in low frequencies can be primarily achieved by transitions from NLoS to LoS for more favorable propagation conditions, which compensate for the adverse impact of longer link lengths at altitudes \cite{azari2021evolution}. However, such findings should be re-visited for THz communication since the communication quality drastically drops with the link length. 
%For instance, in designing an integrated access and backhaul for a UAV BS serving ground nodes, although the widely available THz bandwidth provides Tbps backhaul links, the benefit is only reachable within a short range of communication with the ground THz receiver, which limits the space for efficient and optimal UAV placement. 
For instance, though the THz band can support Tbps links for designing an integrated access and backhaul for a UAV BS, the benefits are only reachable within a short range of communication, which limits the space for efficient and optimal UAV placement. 
Accordingly, a more strategic and opportunistic planning of 3D placements is needed. The new strategic planning could take advantage of accurate sensing, mapping, and localization capabilities of THz to plan a smarter deployment and establish a more favorable communication environment. For instance, high resolution sensing in THz can be exploited to provide predictive control of the serving users mobility. Although sensing in THz is highly accurate, the range of sensing is short. To overcome the short range of THz wave, multi-frequency operation can be helpful. UAVs may exploit mmWave sensing capability to detect major environmental changes in longer range and rely on THz for high-resolution sensing capability to track subtle changes in shorter range. 

\subsubsection{Dense Deployment of UAVs}
UAVs have to be closer to receivers/transmitters in THz and hence a higher density of UAVs may be needed for specific task accomplishment such as monitoring or sensing. However, each UAV can offload/disseminate larger data and hence higher throughput can be achieved by each UAV making a higher sum-rate capacity. %Accordingly, there is a trade-off between density of deployment and average sum-rate capacity per deployed UAV. 
Furthermore,  although molecular absorption and high path loss force the dense deployment of UAVs, it may create considerable LoS interference (get worse under frequent beams misalignment) and handovers. % for some application scenarios. Such trade-offs are important to be studied before large deployment of THz-empowered UAVs.%\footnote{ \mahdi{more intermittent links!}}

\subsubsection{Multi-Functionality Design}
As discussed earlier, THz-empowered UAVs significantly benefit from the integration of various functionalities such as communication, sensing, localization, and computing. However, a co-design of such multi-function system is not an easy task. In addition to the general challenges such as waveform design for integrated sensing and communication \cite{chaccour2021seven}, the development of novel models and powerful algorithms with low-complexity that can be executed on UAVs with limited power and computation capability is a major challenge. Furthermore, the algorithms should take into account the sensitivity of several UAVs applications to delay caused by the integrated design. %Therefore, several trade-offs may exist here between latency, energy consumption, and performance.
\subsubsection{Uncertain Medium Condition} 
Air composition and its dependency on the meteorological condition make the THz channel highly uncertain and unreliable particularly in an uncontrolled outdoor scenario. The molecular absorption is highly dependent on the percentage of water vapor in the air and the level of humidity. Therefore, offline optimal network deployment and design aspects can only be valid under certain circumstances, limiting the benefits. To some extent, an online instantaneous sensing and communication can alleviate such issue.

% \mahdi{in the explanation of Figure 2 (if relevant and fine) explain "what performance should be evaluated considering the unique features of THz-empowered UAV systems".}
\subsection{Trade-offs}
Since THz transmitters require highly directional transmission, they comprise large antenna arrays to have pencil sized beams. UAV's motion and wobbling can lead to frequent beam misalignment and thus desiring frequent handovers. Such frequent handovers can cause delay and hence adversely affect the latency as well as reliability performance.
On one hand, higher antenna gain leads to higher data rate, while on the other, it may unfavorably affect the latency and reliability. Moreover, large antenna arrays also result in high energy consumption due to large number of ADCs and RF chains. 
Further, longer flights are desirable for a meaningful data transfer in THz-empowered UAVs at the cost of increased energy consumption.
To counterbalance the shorter communication range, the UAVs can be densely deployed which, however, may lead to high interference and thus affect the reliability.
As a consequence, the underlying trade-offs between \textit{Latency}, \textit{Data Rate}, \textit{Energy consumption}, and \textit{Reliability} can be established.

\begin{figure}
\centering
\includegraphics[height=\columnwidth]{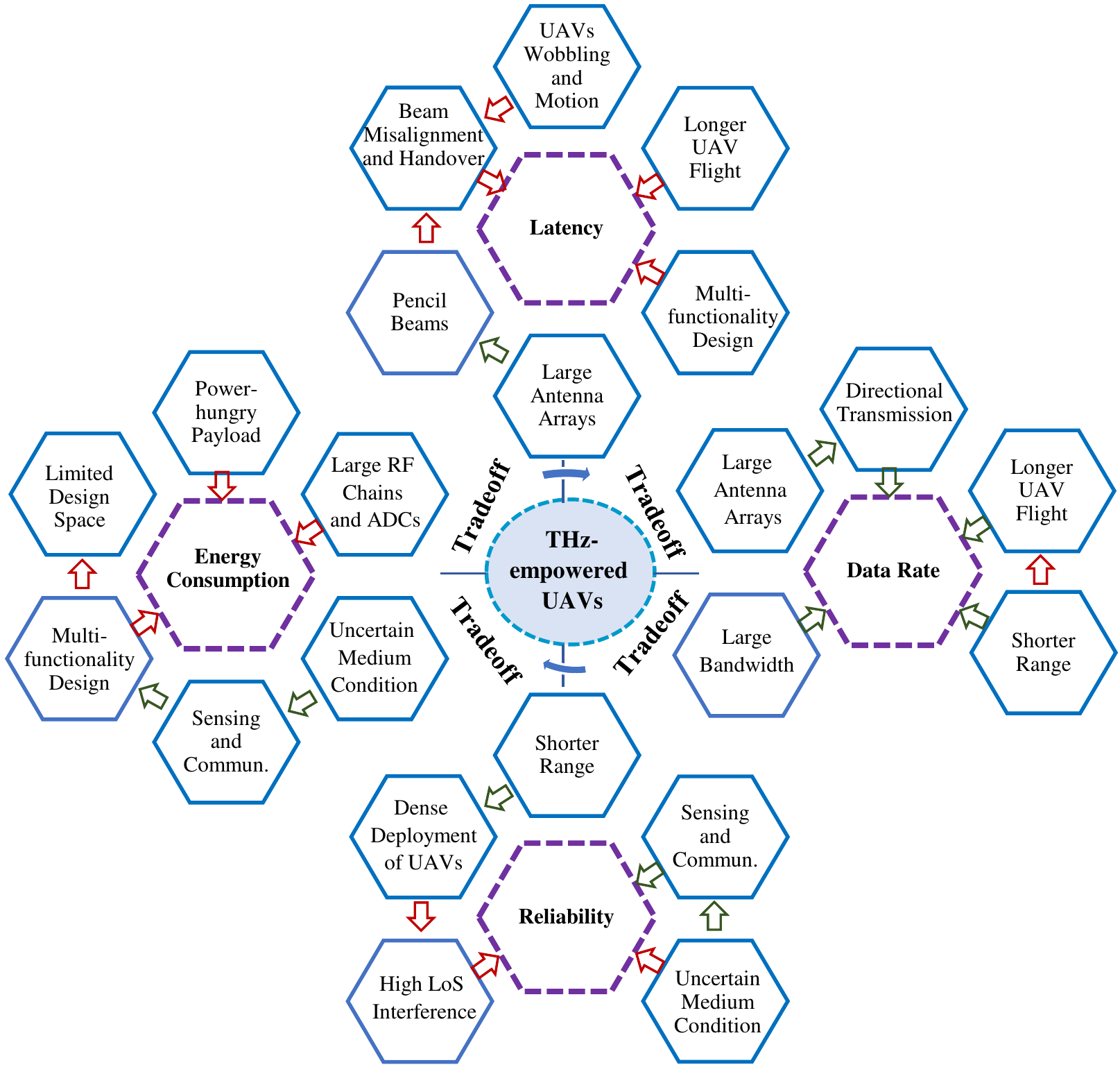}
\caption{Challenges and Trade-offs in THz-empowered UAVs.}
\label{ChaTra}
\end{figure}

Furthermore, integration of various functionalities such as communication, sensing, and localization can improve the reliability and data rate by having a better perception of the environment and beam tracking. However, such multi-functionality design can also lead to an increased payload energy consumption and processing delay. Thus, trade-off between \textit{energy consumption}, and \textit{reliability} exists. In addition, this integrated design may also contribute somewhat to latency as various operations are involved. Hence, the \textit{reliability} can be increased at the expense of \textit{latency} resulting in a trade-off. All these interlinked challenges and trade-offs are also illustrated in Figure~\ref{ChaTra}. Considering the unique features of THz-empowered UAVs, their performance can be characterized by analyzing various performance metrics viz., rate-energy, delay-rate, delay-reliability trade-offs, etc., in addition to the traditional individual key performance indicators (KPIs).
% \subsubsection{Bulky and Heavy Payload}
% Other challenges related to the THz-empowered UAVs include the bulky and heavy payload needed for carrying the related hardware. %\mahdi{Compare with the low frequencies equipment.}

\vspace{-0.1in}
\section{Standardization, Regulations, and Health Issues} \label{sec:regulations}

This section reviews standardization and regulation aspects of UAVs deployment and standardization and health issues related to THz deployment.
\vspace{-0.1in}
\subsection{UAVs Deployment} \label{subsec:UAV_deployment}
Standardization efforts for the UAV deployment started in 2017 when 3GPP in its release-15 studied the LTE-supported UAVs. 
Under this release, it published a technical report TR 36.777 which identified the problems of uplink/downlink interference and provided UE- and network-based solutions for the performance enhancements. In Release 17, the technical specification TS 22.125 identified the UAVs operation requirements via 3GPP system. In the continuation of TS 22.125, TR 22.829 defined new service requirements and KPIs for various 3GPP supported UAV applications. Further, TR 23.754 addressed the mechanisms to support the unmanned aerial systems (UASs) connectivity, identification, and coverage, TR 23.755 identified the application layer support for UAS and provided relevant architectural solutions, and TR 33.854 studied the security aspects of UASs. Technical report TR 22.261 from release-17 and 18 discusses the 5G enhancements for UAV. Specifically, it describes the UAVs application based on the laid out KPIs in TS 22.125 and also plan for the enhanced architecture to support UAS.

On the other hand, international telecommunication union (ITU) has also provided various recommendations to facilitate the UAV based operations and applications. 
For example, ITU-T F.749.10 specified the requirements for communication services of civilian UAVs while ITU-T Y.4421 (ex. Y.UAV.arch) determined the recommendation for functional architecture of UAV controllers using IMT-2020 networks. Further, ITU-T X.677 specified the identification mechanism for UAVs using object identifiers, ITU-T Y.4559 identified the functional architecture and other requirements of base station inspection services using UAVs. Another recommendation ITU-T Q.3060 describes the application of UAVs for fast deployment telecommunication networks in case of any natural disaster.

The IEEE standardization body has also been actively involved in the regulation of the UAV communications and its applications.
For instance, the draft standard IEEE P1936.1 established a framework to support the drone applications where it specifies typical application classes and scenarios and required execution environments. IEEE P1939.1 draft defines a low altitude structure for efficient UAV traffic management. Furthermore, IEEE P1920.1 defines the air-to-air communication for the self-organized aerial adhoc networks where aerial platforms can form a network to communicate with each without requiring the supporting cellular/satellite infrastructure. Another project IEEE P1920.2 identifies the protocol to exchange the information between the vehicles for UASs. 

One of the additional challenges for UAVs deployment is the legislative process. Until a few years ago, the operation of small UAVs was not strictly regulated and they were legally classified as remote controlled toys. But increasing number of UAVs and their potential applications demand for the strict regulation in low-altitude air space to ensure the safe operation and avoid the conflict with manned aircraft. While the large UAVs operating in high-altitude air space can be governed by the same air traffic control (ATC) mechanism as for the conventional aircraft systems, UAVs in low-altitude are difficult to capture by the radar surveillance. Therefore, various countries set specific guidelines that need to be adhered while flying the UAVs to ensure their safe operation. For example, in the USA, Federal Aviation Administration (FAA) and National Aeronautics and Space Administration (NASA) has been jointly framing the concepts of operations for UAV traffic management (UTM) under 400 feet above ground level where ATC services are not available. % \cite{FAA}. 
In Europe, under the project Single European Sky ATM Research (SESAR), U-space programme aims to establish the necessary requirements for integrating UAVs into European airspace. % \cite{SESAR-USpace}.
 As per the instructions from such governing bodies, for example, the UAVs first need to apply for the air space, the flight plan should be submitted before flying, and the pilot should be certified. However, some exceptions can be made to the regulations and they can vary depending on the applications e.g., commercial use, hobbyist flying, research and development purposes etc. 
%\vspace{-0.1in}
\begin{figure*}
     \centering
     \begin{subfigure}[b]{0.7\columnwidth}
         \centering
         \includegraphics[width=\columnwidth]{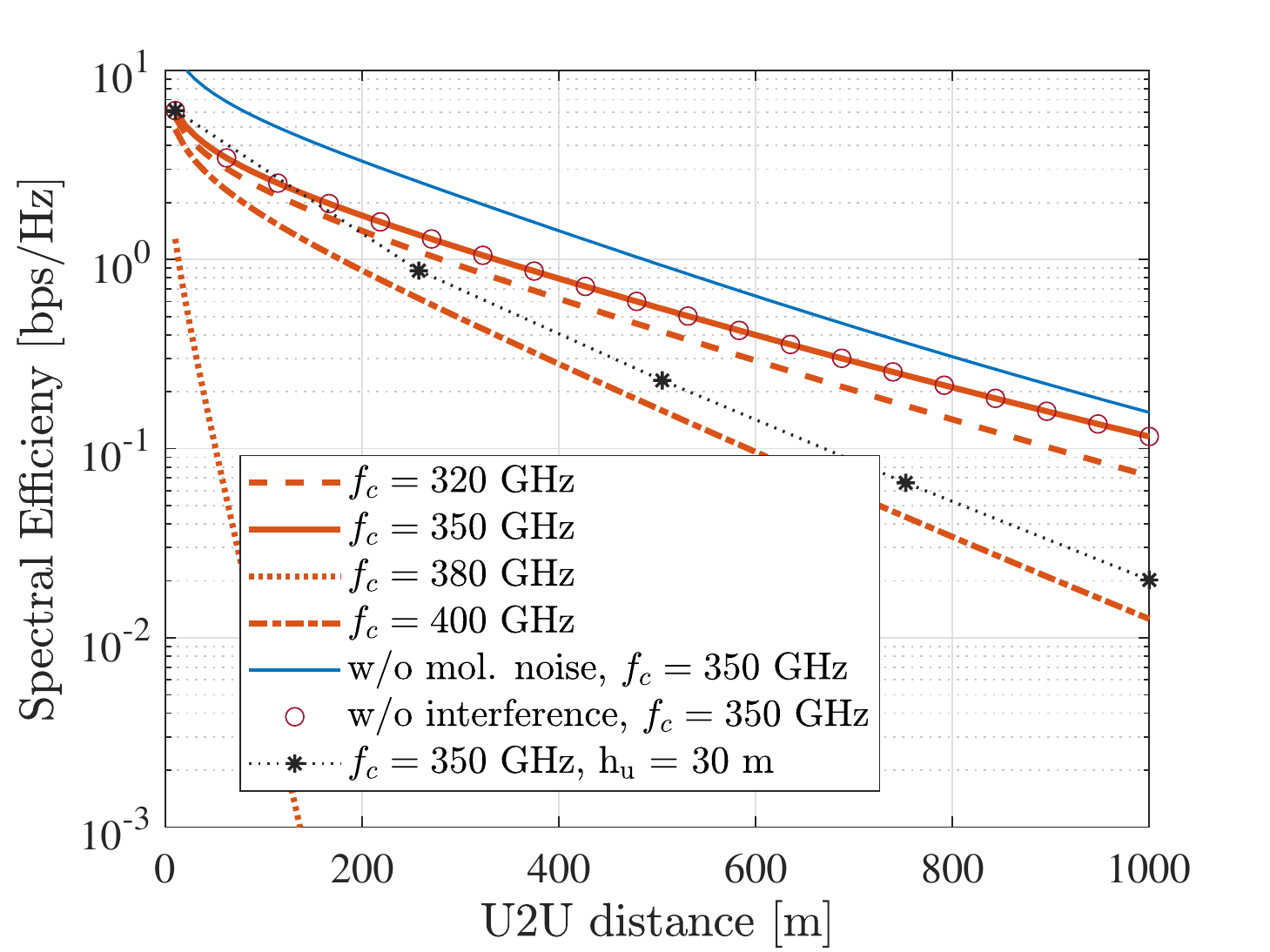}
         \caption{}
         \label{fig:SpEf_dU2U}
     \end{subfigure}
     \hfill
     \hspace{-5mm}
     \begin{subfigure}[b]{0.7\columnwidth}
         \centering
         \includegraphics[width=\columnwidth]{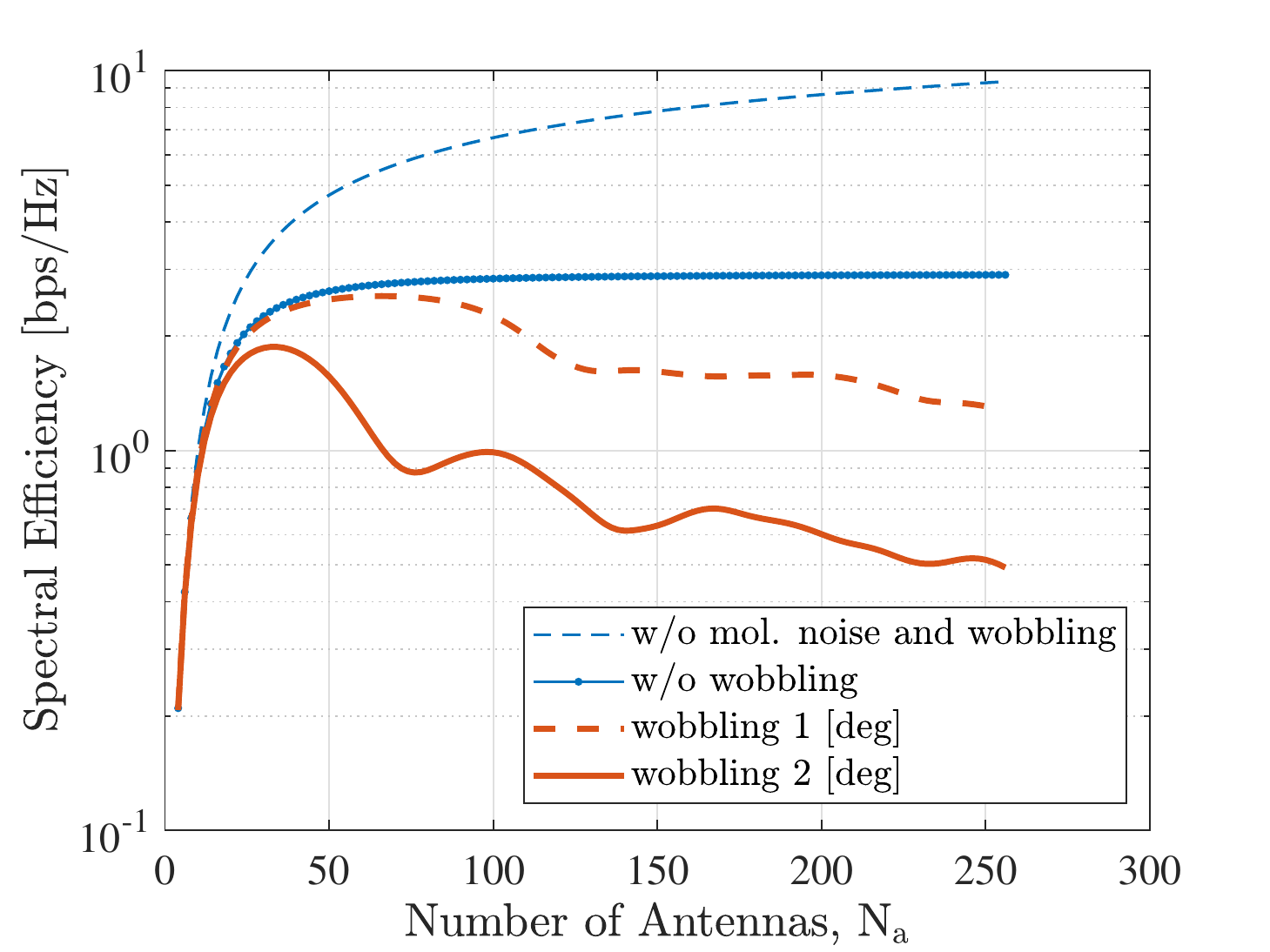}
         \caption{}
         \label{fig:SpEf_numAnt_wobbling}
     \end{subfigure}
     \hfill
     \hspace{-5mm}
     \begin{subfigure}[b]{0.7\columnwidth}
         \centering
         \includegraphics[width=\columnwidth]{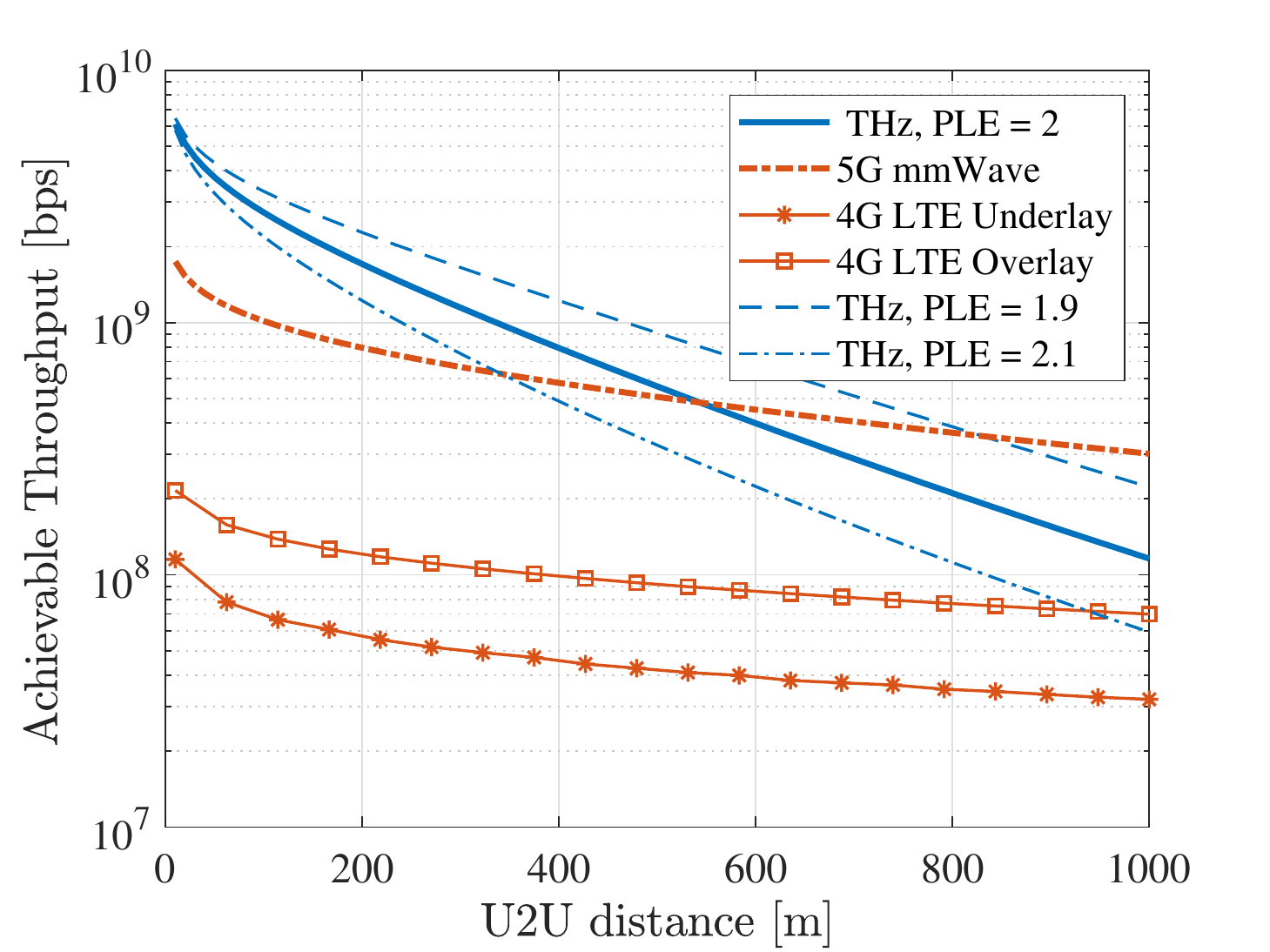}
         \caption{}
         \label{fig:Throughput_dU2U}
     \end{subfigure}
        \caption{(a) Impact of distance, molecular noise, molecular absorption coefficient, frequency of operation, altitude, and interference in U2U THz communication. (b) Number of antennas and UAV wobbling effects where U2U link distance is 100\,m. The wobbling degree shows the maximum variation around the point where the beams are perfectly aligned. (c) THz performance comparison with mmWave and sub-6 GHz for different path loss exponents (PLEs).}%\mahdi{I will add supporting sentences.}}
        \label{fig:three graphs}
\end{figure*}
\subsection{THz Deployment} \label{subsec:THz_deployment}
Although the development of THz communication systems is still in fledgling stage, some standardization and regulation processes have already been initiated. For instance, as an amendment to 802.15.3-2016, IEEE 802.15.3d-2017 is the first standard for THz which aims at alternate PHY layer at frequencies between 252 and 321 GHz. 
%It is developed for switched point-to-point links to achieve the data rates of upto 100 Gb/s. 
Further, the outcomes of world radiocommunication conference (WRC) 2019 provide a regulatory framework for the operation of fixed service (FS) and land mobile service (LMS) applications in frequency bands between 275 GHz and 450 GHz. %\cite{WRC19}. 
%In particular, 275-296 GHz, 306-313 GHz, 318-333 GHz, and 356-450 GHz bands have been identified for use where specific conditions are not necessarily required to protect the passive earth exploration-satellite service (EESS) applications. The other bands 296-306 GHz, 313-318 GHz, and 333-356 GHz can only be used when specific conditions are met to ensure the protection of EESS applications. Whereas, for the entire range of 275-450 GHz, if some radio astronomy (RA) applications exist, certain conditions may be put in place to ensure the protection of RA sites from FS and LMS operations.  
%For the development of nano-network standards at THz frequencies, IEEE P1906.1/Draft 1.0 discusses recommended practices for nano-scale and molecular communication frameworks.
% \begin{figure}
% \centering
% \includegraphics[height=4.0in,width=0.95\columnwidth]{
% Table.pdf}
% \caption{Overview of the Standardization Efforts. }
% \label{fig3}
% \end{figure}

On the flip side, the increasing interest for the THz communications naturally raises the concerns of associated health risks due to high frequency radiation, especially when some non-scientific population still hold the negative perception about the 5G technology, specifically, mmWave and beamforming. In fact, the electromagnetic fields (EMFs) of radio frequencies below $3\times 10^{16}$ GHz is non-ionizing in nature and thus do not possess sufficient energy to ionize the cells and thus avoiding cancer and death risks \cite{samboSurvey15}. However, they do have sufficient energy to push the electrons and ions to higher energy states. The resulting thermal effects may pose certain health issues and therefore, the maximum RF exposure limits are regulated by the regulatory bodies (ICNIRP, FCC, EC, IEEE, etc.) to prevent the harmful heating effects. Nevertheless, it is more important for large antenna arrays in THz to obligate the RF exposure regulations as beamforming can result into elevated power density levels compared with omnidirectional transmissions. Albeit, UAVs supported communications can increase the separation between the THz radiation and the human bodies to reduce the RF exposure to a certain extent.
To date, there is no evident proof of any major health concerns which is supported by scientific community \cite{health1}. However, this aspect requires significant research and experimental studies across various domains to analyze and discover any health related issues which may arise due to THz applications.
\begin{figure}
     \centering
         \includegraphics[width=0.9\columnwidth]{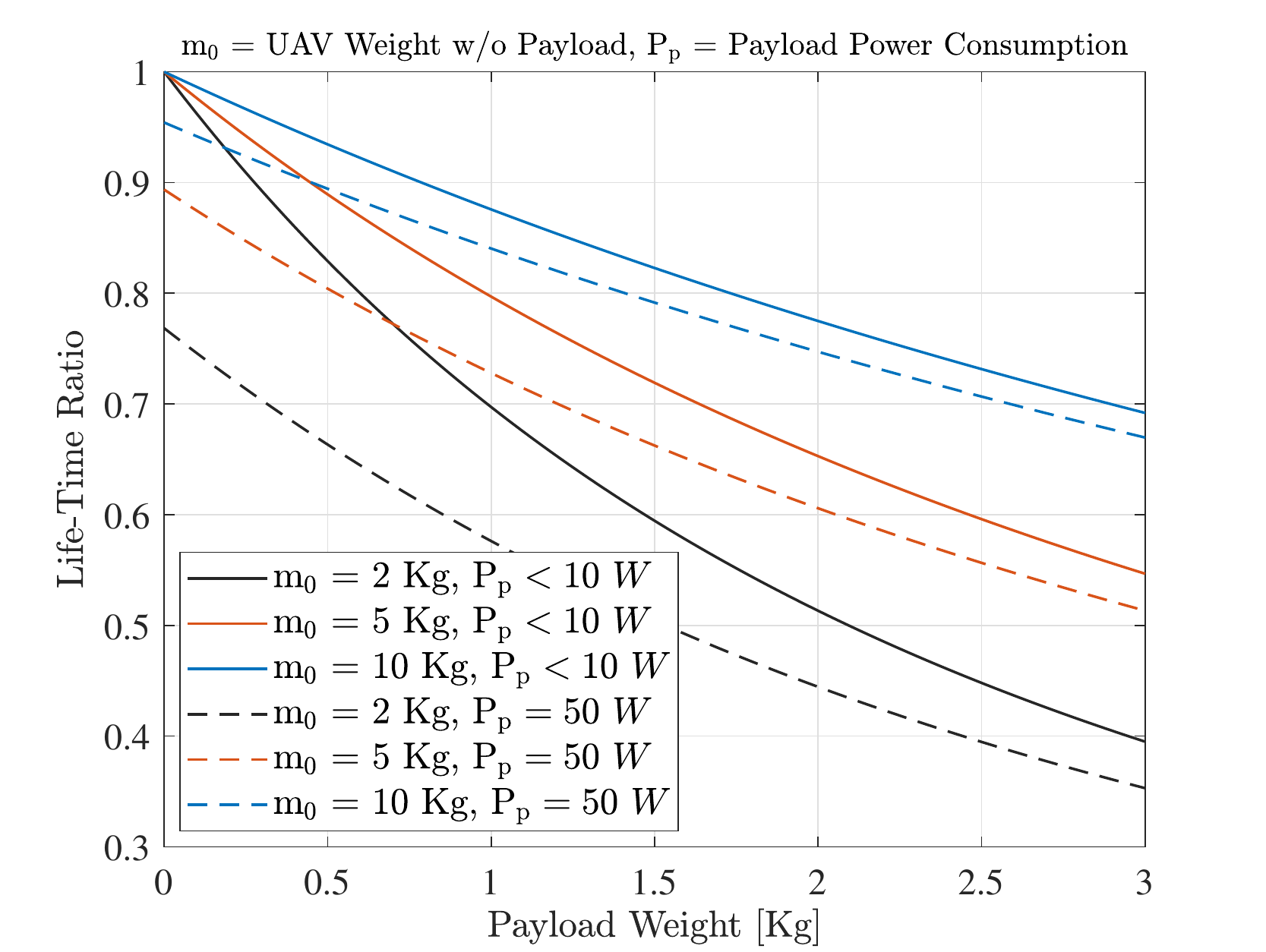}
         \caption{The UAV flying time significantly decreases with the payload related power requirement.}
         \label{fig:EmergyConsumption}
\end{figure}
%\vspace{-0.1in}
\section{Case Study: U2U THz Communication} \label{sec:case_study}

This section provides important insights into various aspects of U2U THz communications. Note that this is an illustrative case study and THz communications are not just limited for U2U.
%Particularly, we numerically study the impact of several key features and challenges mentioned earlier in this paper. 
%\subsubsection{Setup Assumptions}
The simulated scenario includes multiple THz U2U links operating at $\mathrm{f_c} = 350$\,GHz sharing same spectrum. UAVs are distributed according to Poisson point process of density $\lambda_{\mathrm{U}}$. % to account for both random location of UAVs and random number of active UAVs in time. 
To avoid collision and in compliance with regulatory aspects, there is a minimum distance between any two UAVs. %, which is denoted by $\mathrm{r_{exc}}$. 
UAVs are equipped with %uniform linear array (ULA) 
antenna arrays of $\mathrm{N_a}$ elements %UAV transmitter and receiver beams are assumed to be perfectly aligned and hence we ignore the misalignment effect unless otherwise is mentioned. 
and are flying at altitude $\mathrm{h_U} = 100$\,m above a built-up urban area unless otherwise is mentioned. %Other parameters, their definitions, and the default simulation values are listed in Table \ref{table:notation}.    

%Figure \ref{fig:SpEf_dU2U} illustrates the spectral efficiency (SE) of a U2U link operating in THz for various system parameters. 
Figure  \ref{fig:SpEf_dU2U}  illustrates  the  spectral  efficiency  (SE)  of  a  U2U link  operating  in  THz with $\mathrm{N_a}=64$ antenna elements for  various  system  parameters. From this figure, following conclusions are drawn:
\begin{itemize}
    \item THz communication performance significantly drops as the communication distance increases due to severe path loss in THz. At the lower altitude, i.e., $\mathrm{h_u} = 30$\,m, the LoS U2U link is likely to be blocked and hence the performance is worse.%Accordingly, a target SE is only reachable within a limited range of communication. %For example, 1 bps/Hz is reachable up to around 300\,m and 500\,m at  $\{\mathrm{f_c},\mathrm{N_a}\} = \{350,64\}, \{350,256\}$, respectively. 
    \item Molecular noise notably reduces the performance of THz links. This can be observed by comparing the solid blue and solid orange curves corresponding to $\mathrm{f_c}=350$ GHz at 200\,m. As the distance increases, the thermal noise becomes dominant and hence the molecular noise is less involved.
    \item High non-monotonic dependency of molecular absorption to frequency can be drastic and results in an extremely low performance. For example, a peak of molecular absorption at around $\mathrm{f_c} = 380$ GHz results in a remarkably low SE.
    \item Co-channel interference is negligible thanks to the high path loss, molecular absorption loss, and beamforming. %In the figure, the circle markers correspond to a single U2U link without interfering UAVs. 
    As can be seen there is no gap between the circle markers and the solid curve corresponding to $\lambda_{\mathrm{U}} = 0, 10$ interfering UAVs per $\mathrm{km^2}$ respectively. Accordingly, spectrum sharing management could be less complex compared to lower frequencies with large interference impact.
    %\item Increasing the number of antenna elements up to 256 can ensure 1bps/Hz up to an acceptable range of 500m.
\end{itemize}

%Figure \ref{fig:SpEf_numAnt_wobbling} provides important insights regarding the effects of number of antennas and UAV wobbling in U2U THz communication. 
Figure \ref{fig:SpEf_numAnt_wobbling} shows the impact of number of antennas on the SE. In this figure, the wobbling is characterized through variations in the elevation angles of each UAV array, which follows a uniform distribution with the specified maximum degree, i.e., $1^o$ or $2^o$. By comparing the dashed and solid blue curves, it can be seen that after a certain point (e.g., $\mathrm{N_a} = 50$), the slope is much lower in the presence of molecular noise. This means that the increased SE per antenna element is significantly lower which considerably reduces the final performance gain of more concentrated beams in THz. Indeed, the narrower beam and more concentrated energy increases the received power yet it leads to higher re-radiated power by the medium, which acts as molecular noise and degrades the performance. Next, in the presence of UAVs wobbling, increased number of elements and narrower beams could be even detrimental from a certain point onward. This is due to the fact that the wobbling imposes frequent beam misalignment. When the beam is too narrow, the total antenna gain variation with even a little misalignment could be significant resulting in remarkable performance deterioration. Accordingly, there could be an optimal number of UAVs antenna elements depending on the severity of wobbling.

Figure \ref{fig:Throughput_dU2U} compares the achievable throughput of a single U2U link adopting different technologies, i.e., 4G LTE, 5G mmWave, and 6G THz. In the LTE underlay, the uplink spectrum of ground cellular users are re-used by the UAV and, in LTE overlay, the available cellular uplink spectrum is split into two orthogonal portions assigned to ground users and the U2U link. Accordingly, in the former case, ground cellular users generate interference on the UAV, however, in the latter case, there is no ground to air interference at the expense of sacrificing the spectrum. Details of the simulated scenarios for LTE U2U link can be found in \cite{azari2020uav}. This figure reveals that the THz U2U achievable throughput is higher than the mmWave U2U communication up to a certain distance. %For further distances, however, mmWave and LTE may perform better than the THz choice as their performance degradation with increasing distance is slower.  

Figure \ref{fig:EmergyConsumption} illustrates the impact of payload weight and power consumption (due to communication and/or processing) on the overall UAV power consumption and flying time. Apparently, the potential high payload weight for carrying THz-related hardware, e.g., 2.5 Kg for THz spectrometer, significantly reduces the UAV life-time. A non-negligible THz payload power consumption leads to even lower flying time, however, a dedicated payload battery can make up for it. %\mahdi{payload power consumption can be compensated using a second dedicated battery for the payload but the extra payload weight is sth!}

\section{Conclusion}
%Considering the growing interest for the deployment of THz frequencies and significance of aerial networks in the future wireless systems, 
% This paper provided prospective use cases of THz-empowered UAVs covering various functionalities including communication, control, sensing, localization, and imaging. Further, specific design challenges and trade-offs were studied where we noted novel complexities different from the existing technologies \blue{such as mmWave UAV.} We also shed light on the regulatory efforts of UAVs deployment and the health aspects of THz communications, the two barriers which may decelerate commercialization of THz-empowered UAVs. Finally, numerical study on U2U THz communication provided important insights into the impact of various system parameters such as molecular noise, operating frequency, beamforming, and distance on the network performance. 

%In this paper we studied the perspective use cases, design challenges, trade-offs, standardization and regulatory advancements, potential health risks as well as the performance of THz-empowered UAVs. 
Synergies between THz and UAV technologies open the doors for new applications, provide significant performance enhancement, and enable intelligent interactions with the environment. However, the benefits are only reachable if several unique challenges are well addressed. For instance, high payload power consumption and weight of a THz-empowered UAV require new design approaches. Integrated functionalities for enhanced performance needs to be handled onboard with limited computation power. And limited design space leads to very dense deployments or much longer flight time with increased latency and energy consumption.
This paper provided detailed discussions of the opportunities, challenges, and design approaches as well as important factors which determine the benefits of the combination of THz and UAV.

\bibliographystyle{IEEEtran}
\bibliography{main}
% \begin{IEEEbiographynophoto}{\textbf{M. Mahdi Azari}} is a Research Associate with the SnT, University of Luxembourg. 
% \end{IEEEbiographynophoto}
% \begin{IEEEbiographynophoto}{\textbf{Sourabh Solanki}} is a Research Associate with the SnT, University of Luxembourg. 
% \end{IEEEbiographynophoto}
% \begin{IEEEbiographynophoto}{\textbf{Symeon Chatzinotas}} is a Full Professor and Head of the SIGCOM Research Group at SnT, University of Luxembourg.
% \end{IEEEbiographynophoto}
% \begin{IEEEbiographynophoto}{\textbf{Mehdi Bennis}} is an Associate Professor at the Centre for Wireless Communications, University of Oulu and head of the intelligent connectivity and networks/systems group (ICON).
% \end{IEEEbiographynophoto}
\end{document}